\input epsf
\documentclass[12pt,nohyper,notoc]{article} 
\usepackage[dvips]{graphicx}

\def\beq{\begin{equation}}
\def\eeq{\end{equation}}
\def\bea{\begin{eqnarray}}
\def\eea{\end{eqnarray}}

\def \gsim{\mathrel{\vcenter
     {\hbox{$>$}\nointerlineskip\hbox{$\sim$}}}}
\def\gappeq{\mathrel{\rlap {\raise.5ex\hbox{$>$}}
{\lower.5ex\hbox{$\sim$}}}}
\def\lappeq{\mathrel{\rlap{\raise.5ex\hbox{$<$}}
{\lower.5ex\hbox{$\sim$}}}}

\def\Huv{\langle H_u \rangle}

\def\CPV{CP \! \! \! \! \! \! /~~}

\def\mnu{[m_{\nu}]}

\def\nuR{\nu_R}
\def\nuR1{\nu_{R_1}}
\def\e{{\epsilon_1}}

\def\bea{\begin{eqnarray}}   
\def\eea{\end{eqnarray}}

\begin{document}
\vspace*{-1cm}
\renewcommand{\thefootnote}{\fnsymbol{footnote}}
\begin{flushright}
IPPP/03/62 \\
DCPT/03/124 \\
\texttt{hep-ph/0312007} 
\end{flushright}
\vskip 5pt
\begin{center}
{\Large {\bf Leptogenesis and a Jarlskog Invariant
}}
\vskip 25pt
{\bf Sacha Davidson $^{1,}$\footnote{E-mail address: sacha.davidson@durham.ac.uk}
and
Ryuichiro Kitano $^{2,}$\footnote{E-mail address: kitano@ias.edu}
} 
\vskip 10pt  
$^1${\it Dept of Physics, University of Durham,
Durham, DH1 3LE, England} \\
$^2${\it 
School of Natural Sciences, Institute for Advanced Study, Princeton, NJ 08540}
\\
\vskip 20pt
{\bf Abstract}
\end{center}
\begin{quotation}
  {\noindent\small 
The relation between  low energy CP violating phases, and the CP
asymmetry of  leptogenesis $\e$, is investigated.
Although it is known that in general those are independent, there may be
a relation when  a model is specified.
We construct a Jarlskog invariant which is proportional to $\e$ if
the right-handed neutrino masses are hierarchical.
Since the invariant can be expressed in terms of left-handed neutrino
parameters---some measurable, and some not---it is useful in
identifying the limits in which $\e$ is related to MNS phases.
\vskip 10pt
}

\end{quotation}

\vskip 20pt  

\baselineskip 18pt

\setcounter{footnote}{0}
\renewcommand{\thefootnote}{\arabic{footnote}}


\newpage
\section{Introduction}

There has been recently some interest in relating the baryon asymmetry
produced via leptogenesis to leptonic observables. This connection is
not straightforward, because leptogenesis \cite{Fukugita:1986hr} depends
on the masses and couplings of the heavy right-handed (RH) neutrinos,
whereas we can measure parameters of the light left-handed (LH) lepton
doublets. Specifically, in thermal
leptogenesis\cite{Fukugita:1986hr,review,gs}, with hierarchical
$\nu_R$s, one needs the mass and eigenvector of the lightest $\nu_R$ to
compute the baryon asymmetry.

CP violating phases in the leptonic sector could be observed in the
coming years at a neutrino Factory, or possibly in neutrinoless double
$\beta$ decay.  The relation between observable phases and $\e$ has been
explored in many papers, from various perspectives. There is a close
connection in models with 2 right-handed neutrinos \cite{Endoh:2002wm}.
Using invariants, it was shown that there is no ``direct'' relation for
three $\nu_R$ \cite{Branco:2001pq}, in the sense that leptogenesis can
work when there is no observable leptonic CP violation at low energy,
and vice versa. For three hierarchical $\nu_R$s, and hierarchical Dirac
Yukawa coupling constants, there is an analytic approximation that
relates the $\nu_R$ and $\nu_L$ sectors
\cite{Branco:2002kt,Davidson:2002em}.  This was used
\cite{Branco:2002kt} to relate low-energy phases to leptogenesis in
SO(10) models, and used in a bottom-up phenomenological discussion for
SUSY with universal soft masses \cite{Davidson:2002em}.  Many analyses
have been performed in left-right models \cite{LR}, various grand
unified theories \cite{GUTS} and/or textures \cite{textures} or
particular models of the RH mass matrix \cite{SRD} \footnote{
Reconstructing the heavy sector in SUSY \cite{me} and non-SUSY
\cite{AA, Barger} seesaws, and in other $\nu_L$ mass generation scenarios
\cite{AR} has also been discussed, with various assumptions for the
unobserved LH parameters.  }.

The $\CPV$ of leptogenesis is among the $\nu_R$, so is not
easily related to the phases of the left-handed sector.
This paper attempts to circumvent this problem by using a Jarlskog
invariant (closely related to those introduced by Branco Lavoura and
Rebelo \cite{invars}).  For hierarchical $\nu_R$, the CP asymmetry
$\epsilon_1$ produced in $\nu_{R1}$ decay, is the invariant multiplied
by a factor depending on the $\nu_{R1}$ mass and decay rate.  The
invariant is a trace, so summed on all LH and RH indices.  It can
therefore be evaluated in terms of LH or RH parameters.
 The expression in terms of LH parameters is tractable, and identifies
 which phases of the left-handed sector contribute to $\e$.  The
 magnitude of the invariant depends on unknown eigenvalues of the
 neutrino Yukawa matrix.

 Invariants are interesting, in theories with many complex couplings,
 because they better encode the CP violation of physical processes than
 do parametrization-dependent phases.  They also avoid potential
 confusion arising from basis and phases choices.  Since our invariant
 is summed on all indices, it can be evaluated in any parametrization of
 the seesaw, with any basis choice and phase convention. Evaluated using
 its left-handed indices, the invariant depends on the phases of a
 matrix $W = V_L U$, where $U$ is the MNS matrix, and $V_L$ is unknown.
 It is therefore useful for seeing in which models (which $V_L$), the
 MNS phases are important for leptogenesis. In the ``best case'', the
 leptogenesis invariant is proportional to $sin$(the neutrinoless double
 beta decay phase).
%
The relations we find agree with most  previous results. What is
new, is that by using the Jarlskog invariant, we can write the ``$\CPV$
of leptogenesis'' as a sum of terms involving LH phases. This is
model-independent, and only requires that the $\nu_R$ be hierarchical.
The expression for $\e$ in terms of the invariant may also have other
interesting uses, discussed in the second part of section
\ref{sec4}. For instance, the upper bound on $\e$ for degenerate
$\nu_L$, in the (not maximally restrictive) version of \cite{di2}, can
be trivially obtained.

The next section contains notation and more background. The
Branco-Lavoura-Rebelo (BLR) invariant, its relation to $\e$ and various
formulae for the ``leptogenesis invariant'' are in section 3.  In
section 4 we consider what this invariant could be good for, and areas
of parameter space where it can be written in terms of leptonic
observables. Section five  discusses and summarizes the result.

\section{Background and notation}

The Lagrangian for the leptonic sector, in a seesaw \cite{seesaw} model 
 with three right-handed neutrinos,
can be written
\beq
{\cal L} = { Y_e} \bar{e}_R H_d \cdot \ell_L + 
{ Y_\nu}  \bar{\nu}_R H_u \cdot \ell_L + \frac{ M}{2} \nu_R \nu_R + h.c.
\label{Lag}
\eeq
where 
the index order on the Yukawa matrices  is right-left.
We assume there are no SU(2) triplet scalars, whose vevs could
give Majorana masses to the $\nu_L$ directly.

Two relevant bases for the $\nu_R$ vector space are the one where the
mass matrix ${ M}$ is diagonal ($ = { D_M}$), and where the Yukawa
matrix ${ Y_\nu Y_\nu^\dagger}$ is diagonal ($ = { D_Y^2}$). The unitary
matrix ${ V_R}$ transforms between these bases, so in the mass
eigenstate basis \beq { Y_\nu Y_\nu^{\dagger}} = { V_R^\dagger D_Y^2
V_R}~~~.  \eeq

At low energies, well below the $\nu_R$ mass scale, the light (LH)
neutrinos acquire an effective Majorana mass matrix $\mnu$.  In the
vector space of LH leptons, there are three interesting bases--- the one
where the charged lepton Yukawa $ { Y_e^\dagger Y_e} $ is diagonal, the
one where the neutrino Yukawa ${ Y_\nu^\dagger Y_\nu} $ is diagonal, and
the basis where $ \mnu$ is diagonal.  The first (${ D_{Y_e}}$) and last
(${ D_m}$) are phenomenologically important, and are related by the MNS
matrix ${ U} $: $\mnu = {U^* D_m U^\dagger} $ in the ${ D_{Y_e}}$ basis.
The second (${ D_{Y_\nu}}$) relative to the first (${ D_{Y_e}}$)
 can be important for phenomenology in SUSY
models, where ${ Y_\nu^\dagger Y_\nu}
(\equiv V_L^\dagger D_{Y_\nu}^2 V_L$ in the ${ D_{Y_e}}$ basis)
 induces flavour violation via its
appearance in the slepton RGEs \cite{Bor}.  The second and third are
useful to relate LH and RH seesaw parameters\cite{di1}; the matrix $W$
transforms between these bases.  So let us concentrate on these last two
bases for calculating invariants in this and the next section, and
introduce the charged lepton mass eigenstate basis in the discussion,
when relating leptogenesis to low energy leptonic CP violation.

{\it 
In the  basis where ${ Y_\nu}$ is diagonal}, $\mnu$ can be
written
\beq
{ \kappa} ~ \Huv^2 \equiv 
[m_\nu]  =  {D_Y M^{-1} D_Y}  \Huv^2  = { W^* D_\kappa W^\dagger } \Huv^2~~~.
\label{kappa}
\eeq 
The matrix ${ \kappa }$ is convenient, to avoid Higgs vevs
$\Huv$ in formulae. In this paper, ${ \kappa}$ will be in
the ${ D_{Y_\nu}}$ basis, unless otherwise stated.

Twenty-one parameters are required to fully determine the
Lagrangian of eqn (\ref{Lag}). If $Y_e$ is neglected,
only 9 real numbers and 3 phases are required. These can
be chosen in various ways:
\begin{enumerate}
\item `` top-down''---input the $\nu_R$ sector: ${ D_M}$, 
${ D_{Y_\nu Y_\nu^\dagger}},$ and ${ V_R}$.
\item `` bottom-up''---input the $\nu_L$ sector:  ${ D_\kappa}$, 
${ D_{Y_\nu^\dagger Y_\nu}},$ and ${ W}$.
\item ``intermediate''---the Casas-Ibarra parametrization
\cite{meg}: ${ D_M}$, ${ D_\kappa}$, and a complex orthogonal matrix ${ R}
= D_M^{-1/2} Y_\nu D_{\kappa}^{-1/2}$. 
\end{enumerate}
To relate the RH parameters relevant for leptogenesis to the
LH ones, many of which are accessible at low energy, it is useful to consider
the first and second parametrization. 

The baryon asymmetry of the Universe can be generated
in the seesaw model, if enough $\nu_R$s decay out of equilibrium,
and the CP asymmetry in the decay is large enough
\cite{Fukugita:1986hr}. 
Thermal leptogenesis \cite{review,gs}, when
 the lightest right-handed neutrino $\nu_{R1}$
is produced by scattering interactions 
in the plasma, is an  attractive
and comparatively cosmology-independent way for this
to occur. The final baryon asymmetry depends largely  on
three parameters:  
the $\nu_{R1}$ mass $M_1$, its
 decay rate $\Gamma_1$,   and   the
CP asymmetry  $\e$ in the decay.
The decay rate $\Gamma_j$ of $\nu_{Rj}$ can be conveniently parametrized
as 
\beq
\Gamma_j = \frac{[Y_\nu Y_\nu^\dagger]_{jj} M_j}
{8 \pi } \equiv
\frac{\tilde{\kappa}_j M_j^2}{8 \pi}~~,
\label{tildek}
\eeq
 where
$\tilde{\kappa}_j$ is often
of order the elements  of $\kappa$
 (the $\nu_L$ mass matrix), although
it is a rescaled $\nu_R$ decay rate.
The CP violating asymmetry is 
\bea
{\e} & =& \frac{\Gamma_1 - \bar{\Gamma}_1}{\Gamma_1 + \bar{\Gamma}_1} \nonumber \\
     & = & - \frac{1}{8 \pi [Y Y^\dagger]_{11}}  
     \sum_j \Im \{ [YY^\dagger]_{1j}^2 \} g(M_j^2/M_1^2)
\label{epsexact}
\eea
where $\bar{\Gamma}_1$ is 
the decay rate into CP conjugate particles,
 and  $g$ is a kinematic function
which in the MSSM at zero temperature
(see \cite{gs} for more details) reads
\beq
g(x) = \sqrt{x} \left( \frac{2}{x -1} + \ln \left[ 1 + 1/x \right] \right)
\rightarrow  \frac{3}{\sqrt{x}} + \frac{3}{2x^{3/2}}+... ~~.
\label{g}
\eeq
The last limit occurs for  $(M_1/M_j)^2 \ll 1,$  $ j = 2,3$.

\section{Results}

A series of CP invariants for Majorana neutrinos \cite{invars}, is
\beq
 \Im \{ Tr [ M^\dagger Y_\nu Y_\nu^\dagger 
(M M^\dagger)^m M Y_\nu^* Y_\nu^T] \}
\label{III-Branco}
\eeq
for integers $m \geq 1$.
 These can be made  more appropriate for
leptogenesis with a slight modification.  In the $D_Y$ basis, 
\beq
M = V_R D_M V_R^T, ~~~ M^{-1} = V_R^* D_M^{-1} V_R^\dagger
\label{useful}
\eeq
so $M$ can be replaced by $M^{*-1}$ in
the expression (\ref{III-Branco}):
\beq
I_n = \Im \{ Tr [ M^{-1} Y_\nu Y_\nu^\dagger 
M^{-1 *} ( M^{-1} M^{-1 *})^n Y_\nu^* Y_\nu^T] \}~~~.
\label{III-1}
\eeq
It is clear that none of these is the CP asymmetry $\e$ of
thermal leptogenesis, which for sufficiently hierarchical RH neutrinos, can be
written
 \beq
\label{10}
\epsilon_1 \simeq - \frac{3 M_1}{8 \pi  [Y_\nu Y_\nu^\dagger]_{11}} 
\Im \{ [Y_\nu Y_\nu^\dagger]_{1j}^2 \} \frac{1}{M_j}
\eeq
In the remainder of the paper, when we refer to $\e$, we
will mean this approximation, unless otherwise stated.
(A significant hierarchy in the $\nu_R$ is required
for this approximation to be reliable, as discussed in
\cite{AS}.) $\epsilon_1$ is also an invariant, in that it can be evaluated
in any basis---provided the external index ``1'' is  the
lightest RH neutrino $\nu_{R1}$. So it differs from the BLR invariants
(eqn \ref{III-Branco}),
in that one must know the $\nu_{R1}$ mass  and eigenvector
to evaluate it. This complicates  attempts to relate leptogenesis
to low-energy CP violation.  The invariants of eqn (\ref{III-1})
do not suffer this problem; since all indices are summed
in the traces, the invariants can be
evaluated in the LH or RH vector spaces. 
Specifically, $I_1$ can be written in terms of LH quantities as
\bea
I_1 & =& \Im \{ {\rm Tr} [ \kappa^\dagger \kappa \kappa^\dagger (Y_\nu^T Y_\nu^*)^{-1}
 \kappa (Y_\nu^\dagger Y_\nu)^{-1} ] \}  \nonumber \\
&= & \Im \{ {\rm Tr} [ W D_\kappa^3 W^T D_{Y_\nu}^{-2} W^* D_ \kappa W^\dagger 
D_{Y_\nu}^{-2} ] \}~~~.
\label{LHI1}
\eea
There are similar invariants in \cite{invars}, but with
$Y_e$ replacing $Y_\nu^{-1}$. Using $Y_\nu^{-1}$ changes the matrix
transforming from $D_{\kappa} \rightarrow  D_{Y}$, and
the relative weighting of terms in the sum.

\begin{figure}[t]
\begin{center}
\includegraphics[width=10.2cm]{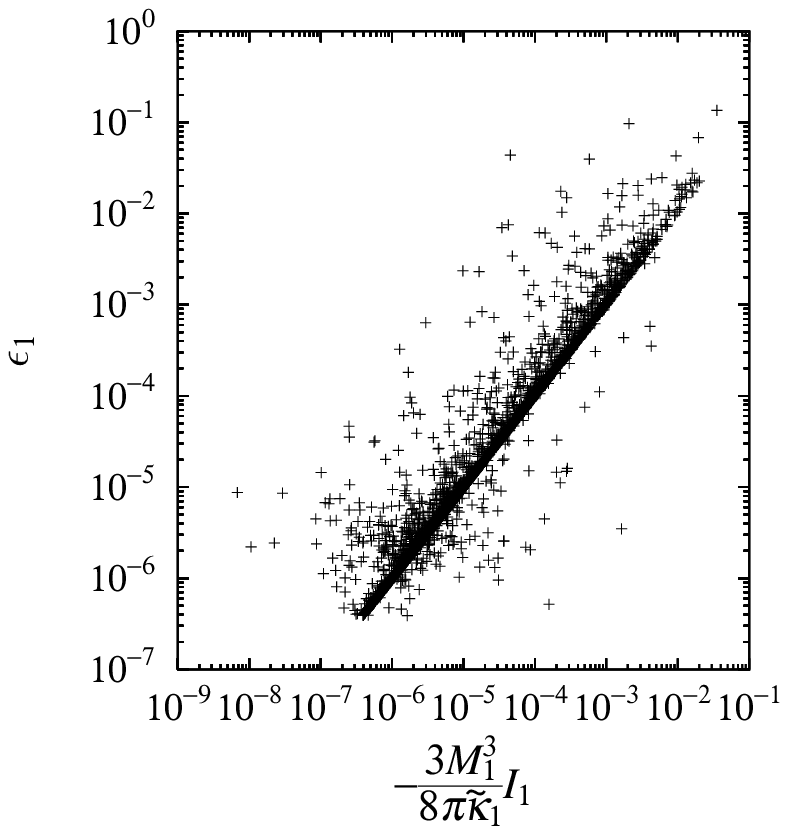} 
\hspace*{-3.7cm}
\includegraphics[width=10.2cm]{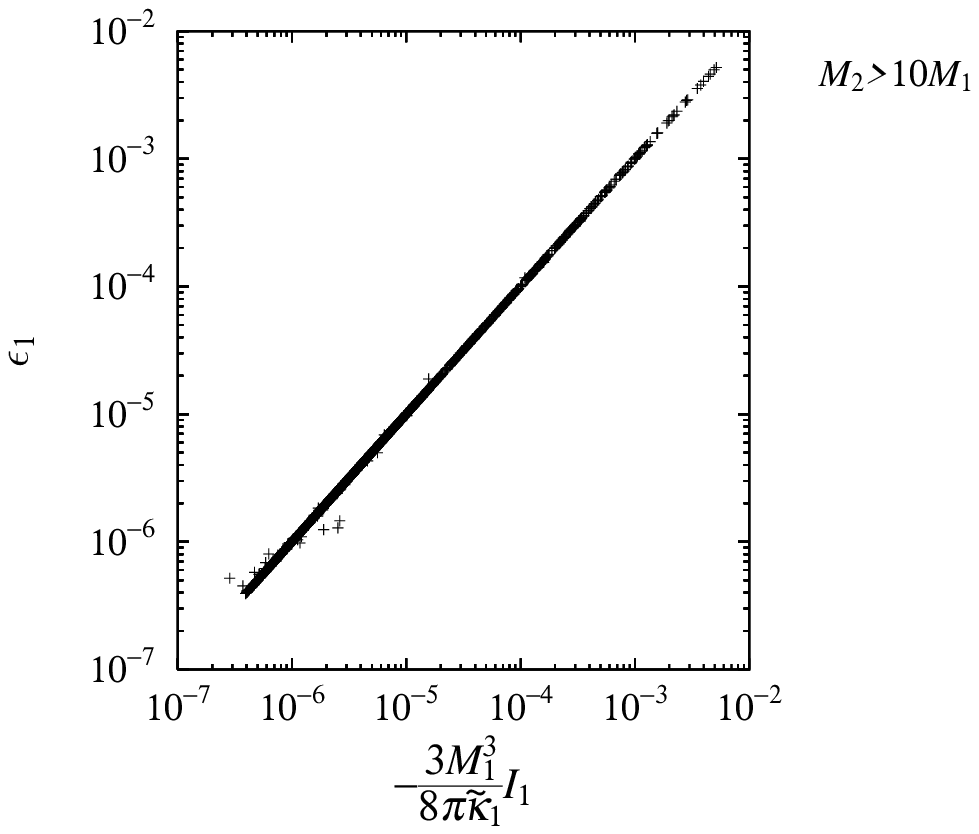} 
\end{center}
\caption{
The scatter plot of $\e$ and $I_1$. We randomly generate the parameters
which reproduce correct neutrino masses, mixing angles and large enough
value of the baryon asymmetry of the universe. In the right figure, we
further require the mild hierarchy of $M_2>10M_1$.
} 
\label{I1_vs_eps}
\end{figure}

In a two generation model, it is clear that 
$I_1 \propto  
 \e$.  
This is not surprising, because the two generation model only has one
phase (if $Y_e$ is neglected). This phase is also parametrized
by the invariant (\ref{III-Branco}), with $m = 1$, as discussed in \cite{AP}.
The difference between $I_1$ and
$\e$ in an arbitrary number of generations is that an RH index is
fixed to ``1'' in $\e$, but summed in $I_n$, which therefore contains
extra terms which are not proportional to $\e$.
However, if the RH neutrino masses are
hierarchical, we can ignore those extra terms  because they
are weighted by
$(M_1/M_j)^{-(1+2n)}$. Explicitly for $n \rightarrow \infty$
\beq
\frac{  M_1^{2+2n} I_n}{  [Y Y^\dagger]_{11} }  \rightarrow  
- \frac{8 \pi }{ 3}  
\epsilon_1\ .
\label{limit}
\eeq
The invariant $I_n$ can be written, in the $\nu_R$
mass eigenstate basis, as
\bea
\frac{M_1^{2n+2} I_n} 
{  [Y_\nu Y_\nu^\dagger ]_{11}}
& =& 
 \frac{1}{
 [Y_\nu Y_\nu^\dagger ]_{11}}
\left[
 \Im \{ [Y Y^\dagger]_{12}^2 \} \frac{M_1}{M_2}
\left( 1 - \frac{M_1^{2n}}{M_2^{2n}} \right)
+ \Im \{ [Y Y^\dagger]_{13}^2 \} \frac{M_1}{M_3}
\left( 1 - \frac{M_1^{2n}}{M_3^{2n}} \right) \right]
 \label{one} \nonumber \\
&& + \frac{1}{
  [Y_\nu Y_\nu^\dagger ]_{11}}
\left(\frac{M_1}{M_2} \right)^{2n+2} 
\left[
\Im \{ [Y Y^\dagger]_{23}^2 \} \frac{M_2 }{ M_3}
\left( 1 - \frac{M_2^{2n}}{M_3^{2n}} \right) 
\right]
\label{term} \\
&\simeq & -\frac{8 \pi}{3} \epsilon_1 + {\rm extra}
\ .
\label{whatn}
\eea
The $\nu_R$ hierarchy must be steep enough, or $n$ large
enough, to ensure that the magnitude of the extra bits
is  much less than  $ 8| \e|$. The second line can be larger than
the $O(M_1^2/M_j^2)$ terms appearing in the first line,
and is bounded above:
\beq
{\rm extra} \leq  \left( \frac{M_1}{M_2} \right)^{2n} 
\frac{\tilde{\kappa}_2}{\tilde{\kappa}_1} M_1 \kappa_3~~~
\label{extra}
\eeq
(where $\tilde{\kappa}_j$ is defined in eqn (\ref{tildek})).
If $\tilde{\kappa}_1$ is small, and $\tilde{\kappa}_2$  is
large, then  $({M_1}/{M_2})^{2n}$ must be small.
Writing $\tilde{\kappa}_2 = |R_{2j}|^2 \kappa_j$, we see
that ${\tilde{\kappa}_2}/{\tilde{\kappa}_1} $ is large
for large imaginary $\theta_{23}$ in $R$, which is
bounded above by requiring that 
$|Y_{33}| = | R_{33} \sqrt{M_3 \kappa_3}|\leq 1$.
This implies ${\tilde{\kappa}_2} \lappeq 1/M_3$. So for
$\tilde{\kappa}_1 \gappeq 10^{-2} \kappa_3$, and $\e \gsim 10^{-6}$ we
obtain that $I_1$ will give a reasonable approximation for $\epsilon_1$
if $({M_1}/{M_2})^{3} < 10^{-7}$.
That is,
\beq
\epsilon_1 \simeq -\frac{3 M_1^3}{8 \pi \tilde{\kappa}_1} I_1 
~~~~{\rm for} ~ 200M_1< M_2~~~.
\label{cdn}
\eeq
   See
the appendix for a more detailed discussion. 
We here draw attention to two points.
First, there are $O(M_1^2/M_j^2)$ corrections to
the  $\e$ of eqn (\ref{10}), similar
to those in parentheses in line (\ref{one}), arising from
the expansion of $g$ in eqn (\ref{epsexact}); see eqn
(\ref{g}). These corrections are always present,
and small if eqn (\ref{cdn}) is satisfied.   
So if we used the invariant $I_2$,  the
second line (eqn (\ref{term})) would be negligible
for a less steep hierarchy, but we would  have to
check that the $M_1^2/M_2^2$ terms  were small enough.
So it is reasonable to work with the simplest invariant,
$I_1$.
Second, in the parameter space where
$\tilde{\kappa}_2 \gg \kappa_3$,
the ``physical'' $\nu_R$ parameters (masses $M_i$ and
Yukawa couplings $Y_{ij}$) are tuned against each other
(recall that in the $\nu_R$ and $\nu_L$ mass eigenstate
bases, $\tilde{\kappa}_2 = [YY^\dagger]_{22}/M_2$ and
$\kappa_2 = [Y^T M^{-1} Y]_{22}$).  These cancellations appear
less unnatural when $M_2 \sim M_3$ (see the model in
the Appendix), so we did
not impose $M_2 \ll M_3$ in deriving the bound (\ref{cdn}).
On the other hand, if we reject such
tuning, then   
$\tilde{\kappa}_2 \sim \kappa_3$ in  eqn (\ref{extra}), and
$I_1$ is a good approximation for
$30 M_1 < M_2$.
We can see this in Fig.\ref{I1_vs_eps} (right), where
we make a parameter space
scan using the milder hierarchy  $10 M_1 < M_2$.

In principle, it is interesting to notice that 
$\e$ can be expressed only in terms  of LH parameters:
if the $\nu_R$ masses  satisfy
eqn (\ref{extra}), then
\bea
\frac{[YY^\dagger]_{11}}{M_1^4} 
&\simeq&  \sum_j \frac{[YY^\dagger]_{jj}}{M_j^4}
=  {\rm Tr} [ M^{*-1}M^{-1}M^{*-1}M^{-1} Y Y^\dagger] \nonumber \\
& =& {\rm Tr} [ \kappa^\dagger \kappa (Y^T Y^*)^{-1} \kappa^\dagger (Y^\dagger Y)^{-1}
\kappa (Y^T Y^*)^{-1} ] \equiv T
\eea
So $ \e \simeq  - 3{I_1}/(8 \pi {T})$. In practise,
$T$ is a complicated function of unmeasured parameters.
It looks recognisable in some limiting cases, such
as $V_L = 1$ with negligeable $y_1^2/y_j^2$,  where
$ T \simeq ( |U_{ej}|^2 m_j^2) |m_{ee}|^2/y_1^6$.

 To summarize: 
we can write $\epsilon_1$ proportional to a Jarlskog invariant.  The
coefficient can be written as a function of the mass and decay rate of
the $\nu_{R1}$, both of which are required elsewhere in thermal
leptogenesis calculations, or as a (more complicated) function of
parameters from the LH sector, such as eigenvalues of $Y_\nu$. The
invariant can be written in terms of of LH parameters, so might tell us
about the relative importance of different LH phases for leptogenesis.


Lets try to evaluate  $I_1$ on its
LH indices. This  invariant  
vanishes for degenerate masses or Yukawa eigenvalues 
(although CP violation is possible among degenerate
Majorana neutrinos \cite{degnu}).  So
$D_Y^{-2}$  in (\ref{LHI1}) can be written
\beq
D_Y^{-2} = \frac{1}{y_3^2} { I} +  \left[ \begin{array}{ccc}
\frac{1}{y_1^2} - \frac{1}{y_3^2}& 0& 0 \\
0 & \frac{1}{y_2^2} - \frac{1}{y_3^2} &0 \\
0& 0 &   0
\end{array} \right]
\label{matap}
\eeq
and it is easy to see that in evaluating $I_1$, 
 $D_Y^{-2}$ can be replaced   by the second matrix on
the RHS. 
This simplifies the formula because it removes a row:
\bea
I_1 
%
& = & \kappa_3 \kappa_2 [\kappa_3^2 - \kappa_2^2]
 \Im  \left\{ \left(  \frac{(y_3^2 - y_1^2)}{y_1^2 y_3^2}   W_{13}  
W_{1 2}^{*}   +
 \frac{(y_3^2 - y_2^2)}{y_2^2 y_3^2}    W_{23}  
W_{2 2}^{*}  \right)^2  \right\}  \nonumber \\
&  & + \kappa_3 \kappa_1 [\kappa_3^2 - \kappa_1^2]
 \Im  \left\{ \left(  \frac{(y_3^2 - y_1^2)}{y_1^2 y_3^2}    W_{13}  
W_{1 1}^{*}   +
 \frac{(y_3^2 - y_2^2)}{y_2^2 y_3^2}    W_{23}  
W_{2 1}^{*}  \right)^2  \right\} \nonumber \\
&  & + \kappa_2 \kappa_1 [\kappa_2^2 - \kappa_1^2]
  \Im  \left\{ \left(  \frac{(y_3^2 - y_1^2)}{y_1^2 y_3^2}   W_{12}  
W_{1 1}^{*}   +
 \frac{(y_3^2 - y_2^2)}{y_2^2 y_3^2}    W_{22}  
W_{2 1}^{*}  \right)^2  \right\}
\label{mess}
\eea
where $W$ is defined in eqn (\ref{W}), and $\kappa$ in
eqn (\ref{kappa}).
This is the main result. In the next section,
we will discuss how to simplify it 
\footnote{The approximation for $\e$ used
in \cite{me} can be obtained from the first,third and fifth
terms of eqn (\ref{mess}).}.
 The formula for $I_1$ is more compact if we
replace $(y_j^{-2} - y_3^{-2}) \rightarrow y_j^{-2}$,
for $j = 1,2$. This amounts to dropping corrections of
order $(y_1/y_3)^2  $ and $(y_2/y_3)^2 $ to terms contributing
to $\e$. If the eigenvalues of $Y_\nu$ are hierarchical, this is an 
insignificant modification.

It is also interesting to express $I_n$ in terms of the matrix $R$:
\beq
I_n = - \Im \left\{ {\rm Tr} \left[ D_{\kappa}^2 R^T D_{M}^{-2n} R \right] 
\right\}
\label{eR}
\eeq
where $D_{\kappa}^2$ can be replaced by $D_{\Delta \kappa^2}$,
similarly to (\ref{matap}).
Taking $n \rightarrow \infty$,
 eqn (\ref{eR}) gives an  expression for $\epsilon_1$ in terms
of $R$
\beq
\e  = \frac{3M_1}{8 \pi \tilde{\kappa}_1} \Im \left\{ 
R_{1j}^2 (\kappa_j^2 - \kappa_3^2) \right\}
\label{epsal}
\eeq
which looks slightly different from the formula in \cite{di2}.

\section{relating $\e$ to MNS phases}
\label{sec4}

Figure \ref{I1_vs_eps} tells us that (\ref{mess}) is an almost exact
formula for $- (8 \pi \tilde{\kappa}_1/3M_1^3) \times \e$.  However, the
magnitude of $\e$ depends on unknown parameters, and eqn (\ref{mess})
for the invariant is not illuminating.  In this section, we first assume
that $\e$ is large enough, and study what $I_1$ can tell us
about low energy CP violation in various cases.  Then we will consider
uses of the alternate expression (\ref{epsal}) for $\e$.

The CP violation required for leptogenesis is encoded in $I_1$. The
expression in terms of LH parameters is simple, but of these parameters,
only two mass differences $\kappa_i^2 - \kappa_j^2$ are measured. We
need to identify a few important terms in the sum, and express them in
terms of potentially observable phases, if we wish to relate the CP
violation required for leptogenesis to phases in the MNS matrix.

The matrix $W$ transforms from the LH  neutrino mass
eigenstate basis to to the basis where $Y_\nu^\dagger Y_\nu$
is diagonal. It can be written as
\beq
W = V_L U
\label{W}
\eeq
where $U$ is the MNS matrix, parametrized as
\bea U= \hat{U}\cdot {\rm diag}(1 ,e^{i\alpha} ,e^{i\beta}) ~~~.
\label{UV}
\eea
 $\alpha$ and $\beta$ are ``Majorana'' phases,
 and $\hat{U}$ has the form of the CKM matrix
\beq
 \label{Vdef} 
\hat{U}= \left[ \begin{array}{ccc}
c_{13}c_{12} & c_{13}s_{12} & s_{13}e^{-i\delta} \\
-c_{23}s_{12}-s_{23}s_{13}c_{12}e^{i\delta} & 
 c_{23}c_{12}-s_{23}s_{13}s_{12}e^{i\delta} & s_{23}c_{13} \\
s_{23}s_{12}-c_{23}s_{13}c_{12}e^{i\delta} & 
 -s_{23}c_{12}-c_{23}s_{13}s_{12}e^{i\delta} &
  c_{23}c_{13}  
\end{array} \right]  ~~~.
\eeq
$W$ is parametrized in the same form as $U$.

The matrix $V_L$ diagonalizes $Y_\nu^\dagger Y_\nu$ in the
charged lepton mass eigenstate basis
\beq
V_L Y_\nu^\dagger Y_\nu V_L^\dagger = D_Y^2 ~~~.
\eeq
In  SUSY models, some information may be available about
the $[V_L]_{31}$ and  $[V_L]_{32}$ elements from
slepton-mediated lepton flavour violation \cite{Bor,meg,ellis_et_al}.

If there is one dominant term in
the invariant, then $I_1$  is proportional to $\sin (2 \xi)$ where $\xi$ is a
combination of the phases of the unitary matrix $W$. 
The combination $\xi$, in various limits, is listed in 
 table \ref{tab1}.  The cases are labeled by the $\nu_L$ mass pattern
and a choice of the form of $V_L$, and then constructed by looking for
conditions such that one of the six terms in eqn (\ref{mess}) dominates
$I_1$. 
In the limits where two terms have the same order of
magnitude, we do not identify a ``phase''.
In the table,
and  all the following, we assume that  phases in $V_L$ and $U$ are
$O(1)$. 
The table tells us that the leptogenesis
invariant can be controlled by MNS phases if
the off-diagonal elements of $V_L$ are ``small''.
In this case,  $W_{ij} \sim U_{ij}$, except
for  $W_{13}$, where  
elements of $V_L$ could make the leading contribution.
There are various interesting limits.

\begin{enumerate}
\item  suppose that $V_L \sim 1$,  so  the 12 and 23 angles in $W$ are
large.
The first, second and fifth terms in
eqn (\ref{mess}) could be important, with relative sizes
$\sim W_{13}^2 : (y_1/y_2)^4 : \frac{\kappa_1 (\kappa_2^2 - \kappa_1^2)}
{\kappa_3 (\kappa_3^2 - \kappa_2^2)}$.  The case
where the first term dominates can be particularly 
interesting:
\beq
\epsilon_1 \simeq
- \frac{3}{8 \pi \tilde{\kappa}_1}  M_1^3 I_1  \simeq 
\frac{3 M_1^3}{8 \pi \tilde{\kappa}_1} 
\frac{\kappa_2 \kappa_3}{y_1^4}[\kappa_3^2 - \kappa_2^2]
 c_{13}^2 s_{12}^2 
 |W_{13}|^2 
 \sin(2 \delta_W - 2\beta + 2\alpha) 
\eeq
where $\delta_W$ is the phase of 
$\hat{W}_{13} \simeq \hat{U}_{13} + V_{L12}/\sqrt{2} + V_{L13}/\sqrt{2}$. 
If any one of the terms contributing to $\hat{W}_{13}$
dominates,
then  the phase of $\hat{W}_{13}$ is the
phase of that term (with our preliminary assumption that
all phases are large).
\begin{enumerate}
\item 
 For instance, if $\theta_{13} \gg \theta_{L13}, \theta_{L12}$,
then $\delta_W \simeq \delta $ is the Dirac phase of
the MNS matrix,
and $  \delta- \beta + \alpha$ is the phase of the
$m_2m_3$ term in  the  
$0 \nu 2 \beta$ decay matrix element.
\item If  $\theta_{L12}$ or $\theta_{L13}$ $\gg \theta_{13}$,
then the phase $\delta_W $ is dominated 
by $\CPV$ from  $V_L$, so $\delta$ is ``subdominant''
for leptogenesis  (because it multiplies
$s_{13}$), and $\alpha - \beta$ multiplies $W_{12}$
so could make  some  ``contribution'' to the 
CP violation required for leptogenesis
(see \cite{Davidson:2002em}
for a more detailed discussion). 
\end{enumerate}
\item Another simplifying limit is to take $\kappa_1 \rightarrow 0$.
If this occurs for fixed $y_1$ ($M_3 \rightarrow \infty$), 
the first two terms of eqn (\ref{mess})
remain.
 This is the case
studied in \cite{Endoh:2002wm}.
\item Now consider the possibility that $V_L \sim U^{-1}$, so $\hat{W} = { I}
+ E$, where the matrix elements of $E$ are small.  This can arise in
(SU(5) texture) models where the mixing in the lepton doublet sector is
large, so the MNS matrix has large angles because $V_L^\dagger \simeq
U$.  These parameters are interesting for leptogenesis, because for
fixed $M_1$, the baryon asymmetry is maximized for $\tilde{\kappa}_1
\gappeq \kappa_2/10$  
(The asymmetry produced in $\nu_{R1}$ decay peaks
for $W$ close to, but not equal to $I$. It rises
as $W \rightarrow 1$,
because less of it is washed out by inverse decay as $\tilde{\kappa}_1$
decreases, and also because
$\e \propto 1/\tilde{\kappa}_1$ increases.
However, in the
hierarchical $\nu_R$ 
approximation of eqn (\ref{10}),
$\e$ drops rapidly to zero at $W = I$.). 
 There is little relation
between the phases of the MNS matrix and $\epsilon_1$ in this case. 
This
can easily be seen in the top-down approach: at the scale $M_1$, $V_L$
has large angles, $W$ has small angles, and the phases of $W$ control
leptogenesis. $U = V_L^\dagger W$, so the complex matrix elements of $U$
will be sums of terms from $V_L$ and $W$.  In particular, the Majorana
phases of $U$ will be those of $W$, plus those of $V_L^\dagger \hat{W}$.
\end{enumerate}

\renewcommand{\arraystretch}{1.5}
\begin{table}
\scriptsize
$
\begin{array}{||l|l||c|c|c|c|c|c||}
\hline
\hline
\nu_L ~{\rm masses}  & V_L \simeq &           1         &     2         &     3         &      4         &     5                &        6          \\
\hline \hline
{\rm hierarchical       }   & I          &  W_{13}^2           & y_1^4/y_2^4   &               &                &  \Delta_{sol}^2 m_s/\Delta_{atm}^3
                                                                                                                                &              \\
m_3 \gg m_2 > m_1 = m_s   
                     &            &  \beta - \delta_{(W)} - \alpha 
                                                        &  \beta - \alpha &              &                &        \alpha         & \\ \cline{2-8}
                     & U^{-1}     &  W_{12}^2 W_{13}^2  & y_1^4 W_{23}^2/y_2^4
                                                                        &   W_{13}^2 m_s/\Delta_{sol}
                                                                                         &                &  W_{12}^2 \Delta_{sol}^2 m_s/\Delta_{atm}^3
                                                                                                                                 &              \\
                    &             &   \beta_W - \delta_W - \alpha_W
                                                       &  \beta_W - \alpha_W & \beta_W - \delta_{W}        
                                                                                         &                &        \alpha_W         & \\ \hline
%
%
 {\rm inverse  ~ hier.}  &     I      &m_s W_{13}^2       & m_s y_1^4/y_2^4  & m_s W_{13}^2 & m_s y_1^4/y_2^4
                                                                                        &\frac{ \Delta_{sol}^2}{  \Delta_{atm}}  & \\
 m_2 > m_1 \gg m_3 = m_s  
                      &            &  -
                                                        & -
                                                                          &      -       &     -          &        \alpha         & \\ \cline{2-8}
                      &    U^{-1}  & 
                                                        & m_s W_{23}^2 y_1^4/y_2^4  
                                                                          & m_s W_{13}^2 &      
                                                                                        &  W_{12}^2 \frac{ \Delta_{sol}^2}{  \Delta_{atm}}  & \\
                      &            &    
                                                        &  \beta_W - \alpha_W 
                                                                          & \beta_W - \delta_W 
                                                                                          &                &        \alpha_W         & \\ \hline
%
%
%
{\rm degenerate}         &  I      &  W_{13}^2       &   y_1^4/y_2^4  &   W_{13}^2 &   y_1^4/y_2^4
                                                                                        &\frac{ \Delta_{sol}^2}{  \Delta_{atm}^2}  & \\
 m_3 \gappeq m_2 \gappeq m_1  
                      &            &  -
                                                        & -
                                                                          &      -       &     -          &        \alpha         & \\ \cline{2-8}
                     &    U^{-1}  & 
                                                        &  W_{23}^2 y_1^4/y_2^4  
                                                                          &  W_{13}^2 &      
                                                                                        &  W_{12}^2 \frac{ \Delta_{sol}^2}{  \Delta^2_{atm}}  & \\
                      &            &   
                                                        &  \beta_W - \alpha_W 
                                                                          & \beta_W - \delta_W 
                                                                                          &               &        \alpha_W         & \\ \hline
\hline
\end{array}
$
\caption{The combination of  phases which contribute to 
 leptogenesis in several limiting cases. The numbers 1--6 represent the
 dominant terms in eqn.(\ref{mess}). Reading along a row, the upper
entries are the relative order of magnitude of the terms. If one
term dominates, then $\e \propto \sin($phase written below) $+$
small corrections. ($\Delta^2_{sol}$ is the solar mass squared difference,
and the phase convention of $W$ is that of eqns (\ref{UV}) and (\ref{Vdef})). 
}
\label{tab1}
\end{table}
\renewcommand{\arraystretch}{1}

We now come to implications of eqn (\ref{epsal}).
In the  hierarchical $\nu_R$ approximation, 
where $g(x)$ is taken as the first term of
eqn (\ref{g}), 
there is an upper bound \cite{HMY,di2} 
 on $\epsilon_1$ :
\beq
\epsilon_1 \leq \frac{3}{8 \pi} M_1 
(\kappa_3 - \kappa_1)
\label{di2bd}
\eeq
 A stronger bound was presented 
in \cite{BdBP}, but see
\cite{AS} for a careful discussion.  Using eqn (\ref{epsal}), 
it is trivial to reproduce  the 
degenerate $\nu_L$ ($m_3>m_2>m_1 \sim m$)   bound    of  \cite{di2}:
\beq
| \e  | \leq \frac{3 M_1}{8 \pi \kappa_1} \sum_j \Im \{ R_{1j}^2 \} 
(\kappa_3^2 - \kappa_j^2)  \leq
\frac{3 M_1 }{16 \pi  {\Huv}^2}
\frac{\Delta^2_{atm}}{ m}
\eeq
where the first inequality follows from  $\tilde{\kappa}_1 \geq \kappa_1$
\cite{di2}. $R$ is a complex orthogonal matrix, so 
 the sum is $\leq \Im \{ R_{13}^2 \} \Delta^2_{atm}/(m
{\Huv}^2)$, and max $ \Im \{R_{13}^2 \} = 1/2$, which gives the second
inequality.

\section{Discussion}

The CP asymmetry $\e$ is the imaginary part of a complex number, so it
is not clear whether there is a ``leptogenesis phase'': to have a phase,
$\e$ would need a uniquely defined real part.  This problem was neatly
addressed in \cite{HMY}, by defining an ``effective'' $\CPV$ parameter
$\delta_{HMY} = \epsilon_1/\epsilon_{1}^{\rm max}$, where
$\epsilon_1^{max}$ is the upper bound on $\epsilon_1$ of eqn
(\ref{di2bd}). $\delta_{HMY}$ has desirable behaviour in two limits; it
goes to zero when there is no $\CPV$, and to 1 for maximal CP
violation. However, one must tune the magnitude of CP conserving
parameters, as well as phases, to obtain $\delta_{HMY} \rightarrow 1$,
so this definition has not accomplished the separation between CP
conserving and CP violating parameters, which speaking about phases
implies. Another approach is to consider specific models, where $\e$ can
be written as a function of real parameters $\times \sin (phases)$.

In $\CPV$ theories which contain many phases, the physical observables
are combinations of complex couplings, which can  be related to
Jarlskog invariants.   The latter are perhaps more
useful than thinking about couplings of fixed magnitude which have
different phases in different parameterizations. So in this spirit, we
look for an leptogenesis invariant, rather than a leptogenesis phase.

The first thing that must always be said, in discussing
potential connections between MNS phases and leptogenesis,
is that there is no linear relation: leptogenesis can work
when there is no $\CPV$ in MNS, and measuring low energy
leptonic phases does not imply that there  is CP violation
available for leptogenesis. This was clearly and elegantly
shown in \cite{Branco:2001pq}. Any relation depends on 
 the choice of ``independent'' phases. If 
 $\epsilon_1$ and $\delta$ are defined to be  ``independent'',
then it follows that they are unrelated \footnote{This 
arises in the Casas-Ibarra seesaw parametrization.}. Or phrased
in terms of invariants:  at least six independent invariants
are needed to determine the phases of the seesaw. If $I_1$ and
the Jarlskog invariant $J \propto \Im {\rm Tr} [Y_e^\dagger Y_e,
m_\nu^\dagger m_\nu]^3 $ are in this set, then they
have been chosen to be independent.
This particular  choice  is disingenuous,
but illustrates the pitfalls of choosing a parametrization,
calculating $\e$, and drawing conclusions about the relations
between leptogenesis and low-energy CP violation. 
Nonetheless, a connection between $\e$ and $\delta$ is
interesting to proponents of a neutrino Factory---
so what can we say?

For hierarchical $\nu_R$s, the baryon asymmetry
produced in thermal leptogenesis
is  proportional to
a Jarlskog invariant $I_1$.
The invariant encodes the CP violation required for
leptogenesis, and  {\it can be written as a function
of $\nu_L$ parameters.} It depends on the phases
of a matrix $W  = V_L U$, where $U$ is the MNS matrix,
and $V_L$ is unknown. So a model for $V_L$ is required
to establish any relation between between $\e$ and
$\delta$.  The coefficient relating $\e$ to $I_1$
can be written as a function of the $\nu_{R1}$ mass
and decay rate, or of  $Y_\nu$ eigenvalues, $\nu_L$ masses
and the matrix $W$.

In some areas of parameter space, one term dominates
in the invariant $I_1$.
When 
these areas correspond to models, where
the relation of $\epsilon_1$ to MNS phases has been discussed,
we reproduce their results.  For instance,
in the ``best case'' scenario,  the leptogenesis
phase is 
$\delta -\beta + \alpha$,  the phase of the 
$m_2m_3$ term of the neutrinoless double beta decay matrix element.
This would arise if the angles of $V_L$ are
smaller than $\theta_{13}$---$e.g.$ if 
$\theta_{13}$  was measured  close to its current bound,  and 
no LFV decays $\ell_j \rightarrow \ell_i \gamma$
were seen (although   superpartners were discovered
to be light),  then
we could be in this scenario. 
Unfortunately, it is the $cosine$ of $\delta -\beta + \alpha$
which appears in  neutrinoless double beta decay,  so 
there would be no  correlation with the sign \cite{BW} of the
baryon asymmetry. 

Also, in the limit where the smallest $\nu_L$
mass goes to zero, the invariant becomes
simpler, losing many of its terms.
 Or if there are large angles in the unknown matrix $V_L$ (the rotation
matrix from the basis where $Y_e$ is diagonal, to where $Y_\nu$ is
diagonal), then no simple relation between the MNS phases and
leptogenesis is expected: the invariant contains many terms to which
many phases contribute, and there are possible cancellations.

\section*{Acknowledgements} 
S.D. thanks
 and Alejandro Ibarra and Jose Valle
for related conversations long ago,
 the astro and particle people of
Valencia for their ever-sunny welcome, and in particular 
Alessandro Strumia
for useful discussions.
The work of R.K. was supported by DOE grant DE-FG02-90ER40542,
and that of S.D. by a PPARC 
Advanced Fellowship.

\section*{Appendix: how steep is the $\nu_R$ hierarchy?}

If we write $Y_\nu = \sqrt{D_M} R \sqrt{D_\kappa}$, where $R$
is a complex orthogonal matrix and  $M_1 <M_2<M_3$ and $\kappa_1<\kappa_2
<\kappa_3$
are positive,  then eqn (\ref{extra}) becomes
\beq
\left( \frac{M_1}{M_2} \right)^{2} \ll
\frac{\tilde{\kappa}_1}{\tilde{\kappa}_2} \frac{8 \pi \e}{ 3 \kappa_3 M_1} =  
\frac{ |R_{1j}|^2 \kappa_j}
{|R_{2k}|^2 \kappa_k} \delta_{HMY}
\label{cdn2}
\eeq
where $\delta_{HMY} \leq 1$, see eqn (\ref{di2bd})
with
$\kappa_1 \ll \kappa_3$. 
We look for parameters that minimize the RHS, to determine
how steep a hierarchy is required between $M_1$ and $M_2$. 
If $\tilde{\kappa}_1$ decreases, this makes the RHS smaller,
but we impose $\tilde{\kappa_1} \gappeq \kappa_3/100 $
for two reasons: 1) the baryon asymmetry
decreases for $\tilde{\kappa}_1$ much below this value, because
not enough $\nu_{R1}$ are produced, and 2),
some tuning is required to get $\tilde{\kappa}_1 < \kappa_2$,
because the MNS angles are large. 
Instead, $\tilde{\kappa}_2$ can be increased,
 $ |R_{2k}|^2 \kappa_k \sim e^{2 \eta} \kappa_3$,
 when
the ``12'' angle  of $R$ is real,  the ``13'' angle zero, and
the  ``23'' angle  acquires a large imaginary
part $\theta_{23} = \rho + i \eta, \eta \gg 1$.
Imposing that the magnitude of  $[Y_\nu]_{33}$  should be
$\leq 1$,  gives an upper bound on $\eta$ :
$$
e^\eta \lappeq \frac{1}{\sqrt{\kappa_3 M_3}}  
$$
so $\tilde{\kappa}_2 \leq 1/M_3$. Thermal leptogenesis
requires $\e  \gappeq 10^{-6}$ \cite{review,gs},
so eqn (\ref{cdn}) will be a good approximation, for $n = 1$,
if 
$$
\left(\frac{M_1}{M_2} \right)^3 \frac{M_2}{M_3} < 10^{-7}~~.
$$
We allow  $M_2 \sim M_3$, with $y_3 \gg y_2$,  because this seems a natural
way to obtain the cancellations implied by $\tilde{\kappa}_2
\gg \kappa_3$. If
$$
{\cal M} = \left[ \begin{array}{cc}
 0 & M \\
M &  0
\end{array} \right]~~, ~~~ Y = \left[ \begin{array}{cc}
 y_2 & 0 \\
0 &  y_3
\end{array} \right]~~~,
$$
then $\kappa_2 = \kappa_3 = y_2y_3/{M}$
and $\tilde{\kappa}_2 = (y_2^2 + y_3^2)/2M$.
(Notice, however, that line (\ref{term})
is  $\propto M_2^2 - M_3^2$, so is zero
for $M_2 = M_3$.)

\end{document}